# Giant Enhancement of Third Harmonic Generation from $Ge_2Sb_2Te_5$ based Fabry-Perot Cavity

Tun Cao[1,*], Kuan Liu[1], Yutao Tang[2], Junhong Deng[2], Kingfai Li[2], and Guixin Li[2,3,*]

[1]School of Optoelectronic Engineering and Instrumentation Science, Dalian University of Technology, Dalian, China, 116024.

[2]Department of Materials Science and Engineering, Southern University of Science and Technology, Shenzhen, China, 518055.

[3]Shenzhen Institute for Quantum Science and Engineering, Southern University of Science and Technology, Shenzhen, China, 518055.

*Email: ligx@sustc.edu.cn & caotun1806@dlut.edu.cn

## Abstract

Third-order harmonic generation (THG) plays a vital role in microscopy, optical communications etc. Conventional methods of obtaining efficient THG in macroscopic crystal is already mature; however, they will finally limit the miniaturization and integration of on-chip laser sources. To date, THG from either photonic crystals or metamaterials provide compact photonic platforms, however selection of materials remains elusive. Herein, we experimentally demonstrate a giant enhancement of THG efficiency from an air/high index $Ge_2Sb_2Te_5$ (GST225) /gold multi-layered Fabry-Perot cavity. At cavity resonant wavelength in near-infrared regime, the efficiency of THG from a 50 nm thick amorphous GST225 planar film is boosted by 422 times compared to that of nonresonant conditions. Interestingly, the THG efficiency has a dramatic decrease of three orders when the structural state of GST225 is transformed from amorphous to crystalline. Our findings have a potential for achieving ultra-compact nonlinear optical source with high efficiency and switchable functionality.

The nonlinear optical processes, such as second- and third-harmonic generation (SHG and THG), can be used to convert low-energy photons into high-energy ones in optical medium[1, 2]. Unlike the SHG effects which only occur in non-centrosymmetric materials under the electric-dipole approximation of the light-matter interaction, the THG effects are usually supported by any material independent of symmetry[3]. The THG is one of the third-order nonlinear optical processes which has wide applications in all-optical switching via Kerr effect, Four-wave mixing and so on[4, 5]. Recently, there has been increasing interests in achieving highly efficient THG in nanophotonic devices. Notably, the improvements of THG efficiency have been obtained by using plasmonic nanocavities[6], surface plasmon resonance[7-11], photonic crystal waveguides[12], magnetic resonance[13-15], and excitonic resonance[16]. The physical mechanism of these strategies is to strongly localize the electromagnetic waves into the vicinity of the nonlinear medium and in turn pronouncedly enhances the efficiency of nonlinear optical processes[17-23]. Although the approaches elucidated above can offer significant enhancement for the THG signals, the quest for the simpler artificial structures and high performance nonlinear optical medium remains a formidable challenge. It is reported that graphene possesses significantly large third-order sheet conductivity $\sigma_s^{(3)}$, resulting in a strong THG effect[24-31] which also have electrically tunability across a broad bandwidth[32-34]. Nevertheless, the integration of graphene layer with large area device can be hard due to its complex nanofabrication processes. In addition, the efficient THG were demonstrated using silicon photonic crystals[12], silicon metasurfaces[35, 36].

On the other hand, phase change materials (PCMs) based on chalcogenide alloys (Ge-Sb-Te) has attracted intense interests in emerging modern photonic applications, ranging from optoelectronic devices capable of multi-level storage, optical display to their integration into nanophotonic systems[37]. Chalcogenide glass possesses an extraordinary portfolio of optical properties[38]. Due to the ultrafast phase transition speed, high cyclability, and excellent thermal stability, the chalcogenide glass is an ideal dielectric for rapidly tunable photonic systems[39, 40]. Chalcogenide glass-based active metasurface is yet another emerging research field, offering effective approaches towards the reconfigurable devices in a variety of fields such as switchable perfect absorber[41, 42], rewritable metasurface lens[43], and beam steering controller[44, 45]etc. Noteworthy, the phase change of the chalcogenide glass from amorphous to crystalline and its backward direction take only about few tenth of nanoseconds[46], which can be experimentally realized by using external thermal stimuli such as heat, electrical or optical pulses[47]. It is expected that the development of a fully addressable, highly integrated phase-change device also benefits the employment of chalcogenide glass for nonlinear optical functionalities.

In this work, we experimentally report the giant THG enhancement from a planar Fabry-Perot (F-P) cavity consisting of tens of nanometer thick $Ge_2Sb_2Te_5$ thin (GST225) film sandwiched by a 100 nm thick gold (Au) reflector and the air. Herein, the low optical losses within the amorphous GST225 dielectric layer in the near-infrared (NIR) regime and the strong cavity resonance result in significant enhancement of THG signals. It is shown that under the cavity resonant condition, the efficiency of THG is 422 times higher than that of the non-resonant condition at the wavelength of λ=1200 nm. The GST225 dielectric has a crystallisation temperature of 433 K ($T_C$ = 433 K) and a melting temperature of 873 K ($T_M$ = 873 K)[43]. By heating the GST225/Au dual-layer structure above the $T_C$ but below the $T_M$, the as-deposited amorphous GST225 can be crystallised. Such a phase transition redshifts the resonant frequency of the Au-GST225-air cavity, therefore, the proposed optical cavity can selectively tune the efficiency of THG light by switching the phase of GST225. Interestingly, we find that the THG response can be hugely enhanced by slightly increasing the thickness of the as-deposited amorphous GST225 thin film, for example, the maximum THG responsivity can be increased from 0.263 to 8.99 as varying the thickness of GST225 spacer of the cavity from 35 nm to 50 nm. The experimental results of spectral resolved THG agree well with the nonlinear optical calculations based on Finite-Difference Time-Domain (FDTD) codes. This work provides a firm basis for comprehension of third-order nonlinear optical process in high index chalcogenide thin film, which paves the way towards future design of chalcogenide glass-based nonlinear optoelectronic devices for a series of applications such as frequency conversion, all-optical switching etc.

## Results

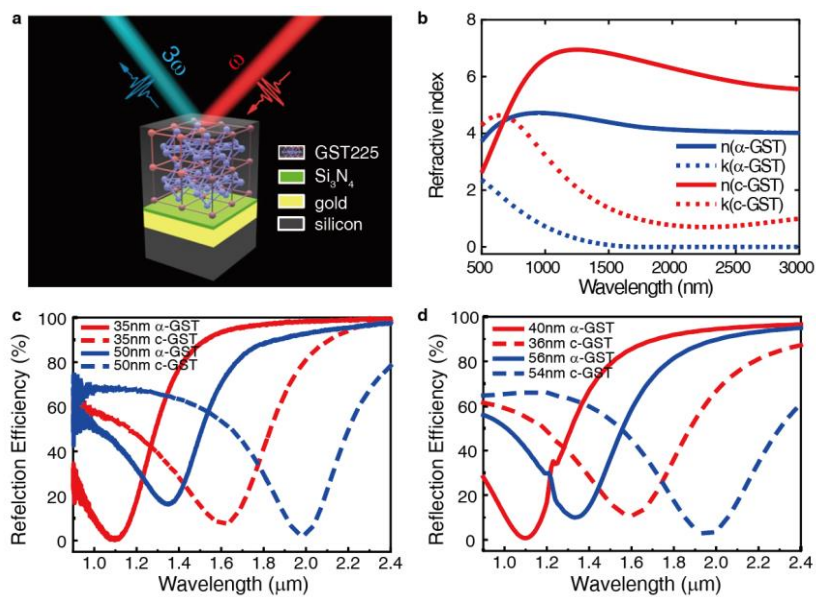

**Figure 1** | (a) Schematic of a planar optical cavity consisting of GST225 spacer over a 5 nm $Si_3N_4$ buffer layer, 100 nm Au bottom mirror, and 500 μm thick Si (001) wafer, where the GST225 layer has the various thicknesses of 35 nm and 50 nm, respectively. (b) The complex refractive index of 50 nm thick GST225 film for both the amorphous and crystalline phases, where the refractive index from 0.5 to 3 μm is obtained using spectroscopic ellipsometry. (c) Experimentally measured, and (d) numerically calculated reflectance spectra for the air-GST225-Au planar cavity. The GST225 films with amorphous (solid lines) and crystalline (dashed lines) states have the different thicknesses of 35nm (red lines) and 50 nm (blue lines), respectively.

In Figure 1(a), we schematically show a planar F-P optical cavity consisting of GST225 spacer, Au and air cladding layer. Firstly, a 100 nm thick Au laminate is deposited on a silicon (Si) substrate using an e-beam evaporator. Afterward, a 5 nm thick $Si_3N_4$ barrier layer is coated onto the Aufilm to forbid interlayer diffusion and interfacial reactions. Then, a GST225 dielectric layer was sputter-deposited on top of the $Si_3N_4$ film (see supplementray information (SI) for details). We fabricate two pieces of dual-layer structures with the different thicknesses of GST225 layers of $t_{GST}$ = 35 and 50 nm, respectively, while fixing the thickness of the Au mirror at $t_{Au}$ = 100 nm.

In Figure 1(b), we measured the complex refractive index of the 50 nm thick planar GST225 film deposited on the Si substrate for both the amorphous (shown in blue lines) and crystalline (shown in red lines) structural phases. The complex refractive indexes of GST225 ($n_{GST}= n + ik$) were obtained via a variable angle spectroscopic ellipsometer. As is seen, both the real and imaginary parts of the index is significantly dispersive over the entire visible-near infrared (VIS-NIR) spectral regions. This vast contrast of $n$ between the two structural phases was achieved by a change to the chemical bonding from covalent in the amorphous state to the resonant bonding after being crystallised[38]. By heating the as-deposited amorphous GST225 (a-GST) layer for 30 min at 433 K on a hot plate in a flowing argon atmosphere, the GST225 layer can be crystallised (c-GST)[48].

The linear optical response of the GST225 based planar F-P optical cavity is experimentally and theoretically presented in Figures.1(c) and (d), respectively. In Figure 1(c), we measure the reflectance under a normal incidence using the Fourier transformation infrared spectrometer (FTIR). The solid and dashed lines respresent the reflectance spectra for the amorphous and crystalline phases respectively, while the red and blue lines present the GST225 films with $t_{GST}$ = 35 and 50 nm, respectively. In the reflectance spectra, deep dips occur around λ= 1100 and 1350 nm for the amorphous state when $t_{GST}$ = 35 and 50 nm, accordingly. It is because the high index a-GST thin film together with Au mirror and air cladding layer forms an optical cavity, and exhibits strong destructive interference effects. The reflectance minimum can be redshifted from 1100 and 1350 nm to 1610 and 1989 nm by changing the phase state of GST225 from amorphous to crystalline. Moreover, we found that a variation of the thickness of the GST225 spacer leads to a remarkable change in the reflectance. For example, the resonant mode has a redshift as increasing the thickness of GST225 dielectric layer for both amorphous and crystalline states. In Figure 1(d), we calculate the reflectance spectrum of the air/GST225/Au cavity using the FDTD method. In the model, we employ the measured refractive index of GST225 as shown in Figure 1(b). The planar cavity is normally illuminated with a plane wave. The thickness of both the GST225 and Au layers were set to values which are close to the measured ones. It is found that the calculations agree well with the measurements. A detailed dsicription of our FDTD model can be found in SI. In Figure S1 of SI, we experimentally measured linear cross-polarized (HV) reflectance specta of air/GST225/Au planar cavities with 35 and 50 nm thick amorphous and crystalline GST225 layer by FTIR. The consistent low reflectance in the visible-NIR regimes indicate that the combined cavities act without any polarization conversion in the linear regime.

**Nonlinear Optical Experiment.** We then performed the THG measurements at room temperature using the optical setup schematically illustrated in Figure 2(a) at reflection mode. A Ti: sapphire femtosecond (fs) laser is used to pump an optical parametric oscillator (OPO), where the fs laser produces a pulse with a wavelength of 0.82 μm, the repetition rate of 80 MHz, and pulse duration of ~200 fs. The OPO idler beam of 20 mW is tunable between 1.2 and 1.6 μm and focused onto the sample after passing through a microscope objective lens (N. A. =0.1). The assessed spot size on the sample is ~30 μm in diameter. The THG signals in transmission direction are collected by an infinity-corrected objective lens (N.A.=0.25) onto an Andor spectrometer (SP500i) with an electromagnatic (EM) CCD detector (see SI for details). For a linearly polarized (H-polarization, the electric field of light is parallel to *x*-axis) fundamental wave (FW) at a wavelength of 1450 nm (Figure 2(b)), the intensity of THG signal with the H-polarization is much stronger than that with vertical polarization (along *y*-axis). In Figure S2 of SI, we study the pumping power dependencies of the THG wave at another two points randomly selected on the sample. All data are identically show the cubic dependence upon the pumping power. For the H(FW)-H(THG) measurement (Figure 2(c)), the intensity of THG at the wavelength of 1450 nm shows a cubic dependence upon the pumping power, which agrees well with theoretical prediction.

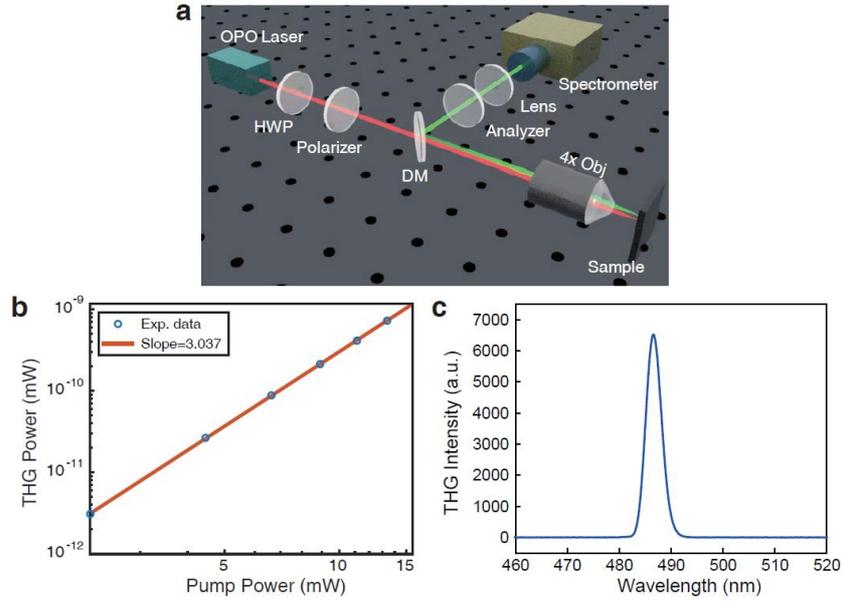

**Figure 2** | (a) Scheme of the experimental set-up for the THG measurment of the air/GST225/Au cavities. (b) The measured intensities of the THG signals under the linearly ploarised FW at λ=1450 nm with the *E*-field parallel to *x*-axis.(c) The spectra of THG at λ= 1450 nm for the H-H measurement.

The spectra of THG responsitivity is dependent with both the thickness and structral state of GST225 film. In Figure 3(a), we show the measured spectra of THG responsitivity for the a-GST device with $t_{GST}$ = 35(red solid line) and 50 nm (blue solid line) . Herein, the THG responsitivity is derivated by $\text{Counts}_{THG}/P_{FW}$, where $P_{FW}$ is pumping power. As is observed, for the a-GST device the THG signal increases with the $t_{GST}$. A giant enhancement of THG response occurs around the wavelength of λ= 1450 nm for $t_{GST}$ = 50 nm. This is because the air/a-GST/Au cavity with $t_{GST}$ = 50 nm resonates at λ= 1350 nm, which can significantly strengthen the light localization around the resonant mode and in turn boost the THG response at λ= 1450 nm.

We then invetigate the effect of the GST225 phase on the THG responsitivity. In Figure 3(b), we measure the spectra of THG responsitivity as transiting the state of GST225 from amorphous to crystalline for $t_{GST}$ = 35 and 50 nm. It is interesting to find that the THG response can be switched off by crystallising the 50 nm thick GST225 spacer. This is because that the GST225 spacer plays a vital role in modulating the resonant condition of air/GST225/Au cavity. The GST225 is an active material that undertakes an amorphous to crystalline phase transition, and its permittivity changes radically during the phase transition, hence tuning the reflectance spectra. As observed in Figures 1(c-d), the spectral position of the resonant reflectance strongly depends on the dielectric environment; changing the phase of GST225 dielectric layer offers a massive shift in its resonant wavelength. For the c-GST cavity with $t_{GST}$ = 50 nm, its resonant wavelength is λ=1989 nm, we can thus not observe the enhancement of THG in the spectral range from 1200 to 1590 nm. Note that, a relatively strong THG responsitivity of 0.75 appears around λ=1580 nm in the c-GST cavity with $t_{GST}$ = 35 nm (see red dashed line in Figure 3(b)), where λ=1580 nm is close to the resonant wavelength of the cavity (λ=1610 nm). Both co-polarized (H-H) and cross-polarized (H-V) THG responsitivites are measured. As is observed, the H-V responses are much weaker than the HH responses (see Figure S3 of SI for details).

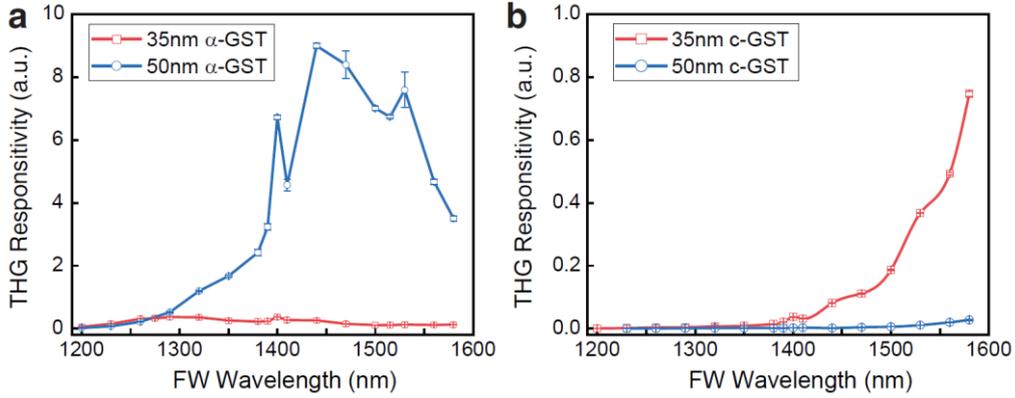

**Figure 3 |** The measured spectral response of THG from the air/GST225/Au F-P cavity with both (a) amorphous and (b) crystalline states for the different thickness of GST225 layer at $t_{GST}$ = 35 and 50 nm, while fixing the thickness of Au reflector at $t_{Au}$ = 100 nm.

**Nonlinear Optical Calculations.** The spectral resolved THG from the GST225 based planar optical cavity is theoretically investigated through the linear optical response of the multilayer system at both the fundamental and third harmonic frequencies. In the nonlinear optical calculation, the relative value of third-order susceptibility $\chi^{(3)}$ of a-GST is assumed to be 1.0 for the FW between 1.2 μm and 1.59 μm while the $\chi^{(3)}$ of Au film is neglected. Firstly, the $E$-field distribution in the GST225 layer is calculated under the excitation of linearly polarized plane wave at the fundamental and THG wavelengths for the cavity having the different thicknesses and phases of the GST225 spacer. Assuming the GST225 layer is homogenous at each local point $r$, the nonlinear polarization can be written as $P_{THG}(r) \sim \chi^{(3)} \cdot \vec{E}(r) \cdot \vec{E}(r) \cdot \vec{E}(r)$. Then, the nonlinear polarization that contributes to the THG can be described by a Green's function $\ddot{G}(r',r)$ at the THG wavelengths, and finally, the complex electric field of THG is given by $\vec{E}_{total}(r') \sim \int \ddot{G}(r',r) \cdot \vec{P}_{THG}(r) d^3 r$. As shown in Figures 4(a) and 4(b), the calculated responsivity of spectrally resolved THG from both a-GST and c-GST based F-P cavity agree well with the measured results. It is found that the calculated

THG responsivity of the a-GST cavity has a maximum value for the fundamental wavelength at 1450 nm. Also, the relative value of $\chi^{(3)} \sim 12$ of c-GST can be extracted by considering both the experimental and calculated results. In Figure S4 of SI, we numerically simulated both co-polarized (HH) and cross-polarized (HV) THG responsivites using FDTD method. The calculated resluts are well consistent with our measured results (see Figure S3 of SI).

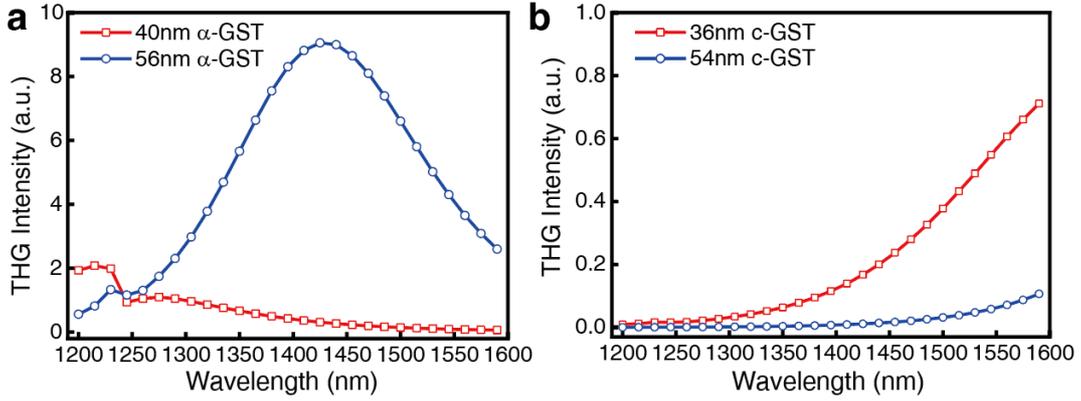

**Figure 4** | The calculated THG spectra of the air/GST225/Au cavity with both (a) amorphous and (b) crystalline states for the different thickness of GST225 layer at $t_{GST}$ = 35 nm, and 50 nm, while fixing the thickness of the Au reflector at $t_{Au}$ = 100 nm.

## Conclusions

In conclusion, we have studied the ultracompact THG platforms by integrating the high index phase change material GST225 into a planar Fabry-Perot optical cavity with Au and air as mirrors. The THG enhancement can be controlled either by tuning the thickness of GST225 layer or its phase states. The giant efficiency switching effect of THG in the proposed devices also promise the great application of tunable THG source or optical modulator in the NIR region mediated by the excellent phase transition performance of GST225. This work paves the way for novel ultrathin chalcogenide film-based nonlinear photonic devices, where the tuning of THG may be explored to fulfill on-chip strategy for optical signal processing and communications.

## Acknowledgments

T.C. acknowledges support from Program for Liaoning Excellent Talents in University (Grant No. LJQ2015021). G. L. would like to thank the financial support from the National Natural Science Foundation of China (No. 11774145), Guangdong Innovative & Entrepreneurial

Research Team Program (2017ZT07C071), Applied Science and Technology Project of Guangdong Science and Technology Department (No. 2017B090918001) and Natural Science Foundation of Shenzhen Innovation Committee (No. JCYJ20170412153113701). The authors acknowledge Advanced Chalcogenide Technologies & Applications Lab from Singapore University of Technology and Design for providing the samples for conducting the research.

# Author contributions

The manuscript was written through contributions of all authors. All authors have given approval to the final version of the manuscript.

# Additional information

**Supplementary Information** accompanies this paper.

**Competing financial interests:** The authors declare no competing financial interests.

# References

1. Kauranen M, Zayats AV. Nonlinear plasmonics. *Nat. Photonics* **6,** 737 (2012).
2. Boyd R. Nonlinear Optics 3rd edn (New York: Academic). (2008).
3. Shen Y-R. The principles of nonlinear optics. *New York, Wiley-Interscience, 1984, 575 p.***,** (1984).
4. Moss DJ, Morandotti R, Gaeta AL, Lipson M. New CMOS-compatible platforms based on silicon nitride and Hydex for nonlinear optics. *Nat. Photonics* **7,** 597 (2013).
5. Renger J, Quidant R, Van Hulst N, Novotny L. Surface-enhanced nonlinear four-wave mixing. *Phys. Rev. Lett.* **104,** 046803 (2010).
6. Cai W, Vasudev AP, Brongersma ML. Electrically controlled nonlinear generation of light with plasmonics. *Science* **333,** 1720-1723 (2011).
7. Tsang TY. Surface-plasmon-enhanced third-harmonic generation in thin silver films. *Opt. Lett.* **21,** 245-247 (1996).
8. Lapine M, Shadrivov IV, Kivshar YS. Colloquium: nonlinear metamaterials. *Rev. Mod. Phys.* **86,** 1093 (2014).
9. Li G, Zhang S, Zentgraf T. Nonlinear photonic metasurfaces. *Nat. Rev. Mater.* **2,** 17010 (2017).
10. Chen S, Li G, Zeuner F, Wong WH, Pun EYB, Zentgraf T*, et al.* Symmetry-selective third-harmonic generation from plasmonic metacrystals. *Phys. Rev. Lett.* **113,** 033901 (2014).
11. Li G, Chen S, Pholchai N, Reineke B, Wong PWH, Pun EYB*, et al.* Continuous control of the nonlinearity phase for harmonic generations. *Nat. Mater.* **14,** 607 (2015).

12. Corcoran B, Monat C, Grillet C, Moss DJ, Eggleton BJ, White TP, *et al.* Green light emission in silicon through slow-light enhanced third-harmonic generation in photonic-crystal waveguides. *Nat. Photonics* **3,** 206 (2009).
13. Shcherbakov MR, Neshev DN, Hopkins B, Shorokhov AS, Staude I, Melik-Gaykazyan EV*, et al.* Enhanced third-harmonic generation in silicon nanoparticles driven by magnetic response. *Nano Lett.* **14,** 6488-6492 (2014).
14. Shorokhov AS, Melik-Gaykazyan EV, Smirnova DA, Hopkins B, Chong KE, Choi D-Y*, et al.* Multifold enhancement of third-harmonic generation in dielectric nanoparticles driven by magnetic Fano resonances. *Nano Lett.* **16,** 4857-4861 (2016).
15. Melik-Gaykazyan EV, Shcherbakov MR, Shorokhov AS, Staude I, Brener I, Neshev DN*, et al.* Third-harmonic generation from Mie-type resonances of isolated all-dielectric nanoparticles. *Philos. Trans. Royal Soc. A* **375,** 20160281 (2017).
16. Chen R, Lin D, Mendoza B. Enhancement of the third-order nonlinear optical susceptibility in Si quantum wires. *Phys. Rev. B* **48,** 11879 (1993).
17. Zayats AV, Smolyaninov II, Maradudin AA. Nano-optics of surface plasmon polaritons. *Phys. Rep.* **408,** 131-314 (2005).
18. Novotny L, Van Hulst N. Antennas for light. *Nat. Photonics* **5,** 83 (2011).
19. Hanke T, Krauss G, Träutlein D, Wild B, Bratschitsch R, Leitenstorfer A. Efficient nonlinear light emission of single gold optical antennas driven by few-cycle near-infrared pulses. *Phys. Rev. Lett.* **103,** 257404 (2009).
20. Aouani H, Rahmani M, Navarro-Cía M, Maier SA. Third-harmonic-upconversion enhancement from a single semiconductor nanoparticle coupled to a plasmonic antenna. *Nat. Nanotechnol.* **9,** 290 (2014).
21. Grinblat G, Li Y, Nielsen MP, Oulton RF, Maier SA. Enhanced third harmonic generation in single germanium nanodisks excited at the anapole mode. *Nano Lett.* **16,** 4635-4640 (2016).
22. Grinblat G, Li Y, Nielsen MP, Oulton RF, Maier SA. Efficient third harmonic generation and nonlinear subwavelength imaging at a higher-order anapole mode in a single germanium nanodisk. *ACS nano* **11,** 953-960 (2016).
23. Shibanuma T, Grinblat G, Albella P, Maier SA. Efficient third harmonic generation from metal–dielectric hybrid nanoantennas. *Nano Lett.* **17,** 2647-2651 (2017).
24. Cheng J, Vermeulen N, Sipe J. Third order optical nonlinearity of graphene. *New J. Phys.* **16,** 053014 (2014).
25. Cheng J, Vermeulen N, Sipe J. Numerical study of the optical nonlinearity of doped and gapped graphene: From weak to strong field excitation. *Phys. Rev. B* **92,** 235307 (2015).
26. Mikhailov SA. Quantum theory of the third-order nonlinear electrodynamic effects of graphene. *Phys. Rev. B* **93,** 085403 (2016).
27. Hendry E, Hale PJ, Moger J, Savchenko A, Mikhailov S. Coherent nonlinear optical response of graphene. *Phys. Rev. Lett.* **105,** 097401 (2010).
28. Zhang H, Virally S, Bao Q, Ping LK, Massar S, Godbout N*, et al.* Z-scan measurement of the nonlinear refractive index of graphene. *Opt. Lett.* **37,** 1856-1858 (2012).
29. Miao L, Jiang Y, Lu S, Shi B, Zhao C, Zhang H*, et al.* Broadband ultrafast nonlinear optical response of few-layers graphene: toward the mid-infrared regime. *Photonics Res.* **3,** 214-219 (2015).
30. Dremetsika E, Dlubak B, Gorza S-P, Ciret C, Martin M-B, Hofmann S*, et al.* Measuring the nonlinear refractive index of graphene using the optical Kerr effect method. *Opt. Lett.* **41,** 3281-3284 (2016).
31. Hong S-Y, Dadap JI, Petrone N, Yeh P-C, Hone J, Osgood Jr RM. Optical third-harmonic generation in graphene. *Phys. Rev. X* **3,** 021014 (2013).

32. ACS nanoAlexander K, Savostianova NA, Mikhailov SA, Kuyken B, Van Thourhout D. Electrically tunable optical nonlinearities in graphene-covered SiN waveguides characterized by four-wave mixing. *ACS Photonics* **4,** 3039-3044 (2017).
33. Soavi G, Wang G, Rostami H, Purdie DG, De Fazio D, Ma T, *et al.* Broadband, electrically tunable third-harmonic generation in graphene. *Nat. Nanotechnol.*, 1 (2018).
34. Jiang T, Huang D, Cheng J, Fan X, Zhang Z, Shan Y, *et al.* Gate tunable third-order nonlinear optical response of massless Dirac fermions in graphene. *B. Am. Phys. Soc.*, (2018).
35. Yang Y, Wang W, Boulesbaa A, Kravchenko II, Briggs DP, Puretzky A, *et al.* Nonlinear Fano-resonant dielectric metasurfaces. *Nano Lett.* **15,** 7388-7393 (2015).
36. Shcherbakov MR, Shorokhov AS, Neshev DN, Hopkins B, Staude I, Melik-Gaykazyan EV, *et al.* Nonlinear interference and tailorable third-harmonic generation from dielectric oligomers. *ACS Photonics* **2,** 578-582 (2015).
37. Wuttig M, Bhaskaran H, Taubner T. Phase-change materials for non-volatile photonic applications. *Nat. Photonics* **11,** 465 (2017).
38. Shportko K, Kremers S, Woda M, Lencer D, Robertson J, Wuttig M. Resonant bonding in crystalline phase-change materials. *Nat. Mater.* **7,** 653 (2008).
39. Li P, Yang X, Maß TW, Hanss J, Lewin M, Michel A-KU, *et al.* Reversible optical switching of highly confined phonon–polaritons with an ultrathin phase-change material. *Nat. Mater.* **15,** 870 (2016).
40. Cao T, Long T, Liang H, Qin K-R. Reconfigurable, graphene-coated, chalcogenide nanowires with a sub-10-nm enantioselective sorting capability. *Microsyst. Nanoeng.* **4,** 1-8 (2018).
41. Tittl A, Michel AKU, Schäferling M, Yin X, Gholipour B, Cui L, *et al.* A switchable mid‐infrared plasmonic perfect absorber with multispectral thermal imaging capability. *Adv. Mater.* **27,** 4597-4603 (2015).
42. Cao T, Zhang L, Simpson RE, Cryan MJ. Mid-infrared tunable polarization-independent perfect absorber using a phase-change metamaterial. *J. Opt. Soc. Am. B* **30,** 1580-1585 (2013).
43. Wang Q, Rogers ET, Gholipour B, Wang C-M, Yuan G, Teng J, *et al.* Optically reconfigurable metasurfaces and photonic devices based on phase change materials. *Nat. Photonics* **10,** 60 (2016).
44. Cao T, Zheng G, Wang S, Wei C. Ultrafast beam steering using gradient Au-Ge 2 Sb 2 Te 5-Au plasmonic resonators. *Opt. Express* **23,** 18029-18039 (2015).
45. de Galarreta CR, Alexeev AM, Au YY, Lopez‐Garcia M, Klemm M, Cryan M, *et al.* Nonvolatile Reconfigurable Phase‐Change Metadevices for Beam Steering in the Near Infrared. *Adv. Funct. Mater.* **28,** 1704993 (2018).
46. Bruns G, Merkelbach P, Schlockermann C, Salinga M, Wuttig M, Happ T, *et al.* Nanosecond switching in GeTe phase change memory cells. *Appl. Phys. Lett.* **95,** 043108 (2009).
47. Raoux S, Ielmini D, Wuttig M, Karpov I. Phase change materials. *MRS Bulletin* **37,** 118-123 (2012).
48. Lu H, Thelander E, Gerlach JW, Decker U, Zhu B, Rauschenbach B. Single Pulse Laser‐Induced Phase Transitions of PLD‐Deposited Ge2Sb2Te5 Films. *Adv. Funct. Mater.* **23,** 3621-3627 (2013).